\documentclass[12pt,epsf]{article}
\usepackage{epsfig}
\usepackage{amsmath}
\usepackage{amssymb}
\usepackage[nosort]{cite}
\newcommand{\eps}{\epsilon}

\newcommand{\CC}{{\cal C}}

\newcommand{\OO}{{\cal O}}
\newcommand{\HH}{{\cal H}}

\newcommand{\NN}{{\cal N}}

\newcommand{\pint}{\makebox[0pt][l]{\hspace{3.4pt}$-$}\int}

\newcommand{\be}{\begin{equation}}
\newcommand{\ee}{\end{equation}}
\newcommand{\ben}{\begin{eqnarray}\displaystyle}
\newcommand{\een}{\end{eqnarray}}

\newcommand{\vareps}{\varepsilon}

\newcommand{\s}{\sigma}

\newcommand{\la}{\lambda}

\newcommand{\ve}{\varepsilon}
\newcommand{\sectiono}[1]{\section{#1}\setcounter{equation}{0}}

\begin{document}

{}~ \hfill\vbox{\hbox{hep-th/0405243}
\hbox{UUITP-16-04}\break
\hbox{CTP-MIT-3496}}\break

\vskip 1.cm

\centerline{\large \bf Higher Loops Beyond the $SU(2)$ Sector}
\vspace*{5.0ex}

\centerline{\large \rm Joseph~A.~Minahan}

\vspace*{2.5ex}
\centerline{\large \it Department of Theoretical Physics}
\centerline{\large \it Box 803, SE-751 08 Uppsala, Sweden}
\vspace*{2.5ex}

\centerline{and}

\vspace*{2.5ex}
\centerline{\large \it Center for Theoretical Physics}
\centerline{\large \it Massachusetts Institute of Technology}
\centerline{\large \it Cambridge, MA 02139, USA}
\vspace*{2.5ex}

\centerline{\tt joseph.minahan@teorfys.uu.se}

\vspace*{1.0ex}
\medskip
\bigskip\bigskip
\centerline {\bf\large Abstract}

\bigskip\bigskip
We consider the case of coherent gauge invariant operators in
the $SU(3)$ and $SO(4)$ sectors.  We argue that in many cases, these sectors
can be closed in the thermodynamic limit, even at higher loops.
We then use a modification of the Bethe equations which is a natural
generalization of a proposal put forward by Serban and Staudacher
to make gauge theory predictions for the anomalous
dimensions for a certain class of operators in each sector.  We show
that the predictions are consistent with semiclassical string predictions
at two loops but in general fail to agree at three loops.  Interestingly,
in both cases there is one point in the configuration space where the
gauge theory and string theory predictions agree. In the $SU(3)$ case
it corresponds to a circular string with $R$-charge assignment $(J,J,J)$.

\bigskip

\vfill \eject
\baselineskip=17pt

\sectiono{Introduction}

There have many recent successes in comparing the anomalous dimensions
of long coherent single trace gauge invariant operators in $\NN=4$ SYM
with the energies of semiclassical string states (see \cite{Tseytlin:2003ii} 
for
a comprehensive list).  The key ingredients that make such
comparisons possible are the semiclassical string picture
\cite{Gubser:2002tv} and  integrability.  
At the one-loop level it was shown that
the $\NN=4$ SYM dilatation operator for the $SO(6)$ sector can be mapped to
an integrable spin-chain \cite{Minahan:2002ve}.  
In \cite{Beisert:2003jj,Beisert:2003yb} this was extended to
the full one-loop dilatation operator, unifying the results in 
\cite{Minahan:2002ve}
with earlier results from QCD \cite{Lipatov:1993yb,Braun:1998id}.

With integrability, the computation of the anomalous dimension is
reduced to solving a set of Bethe equations.  For long coherent operators,
these equations become one or more integral equations.
These integral equations were first solved for states that are dual to
spinning and circular strings \cite{Beisert:2003xu,Beisert:2003ea}, 
and it was shown that these
solutions were consistent with the one-loop 
semiclassical predictions in
\cite{Frolov:2003qc,Frolov:2003xy,Frolov:2002av}.  
Such strings and their dual operators live in an $SU(2)$
subsector of the full $SU(2,2|4)$ algebra for the dilatation operator.
It was subsequently shown that {\it all} one-loop anomalous dimensions
are consistent with the semiclassical string predictions in this sector,
either by comparing all solutions \cite{Kazakov:2004qf}, or by 
showing that the string sigma model action reduces to the Heisenberg magnet
effective action in the limit where the gauge coupling is taken to
zero \cite{Kruczenski:2003gt}.

It is also possible to make higher loop comparisons. 
Once one
goes beyond one-loop, integrability requires including all orders in
the perturbative Yang-Mills expansion.
Nevertheless, it was first shown that 
the dilatation operator is consistent with integrability at the two-loop
level within a closed $SU(2)$ sector of the full $SU(2,2|4)$ symmetry algebra
\cite{Beisert:2003tq}.  
These authors went further, assuming that integrability was
also present at three loops.  This allowed them to make a conjecture
for the three loop correction to the anomalous dimension for the Konishi
multiplet which has been recently verified by 
\cite{Kotikov:2004er} where they applied an explicit calculation in
\cite{Moch:2004pa} to $\NN=4$ SYM.
It was later shown by Beisert that integrability is consistent with
three loops for the larger closed $SU(2|3)$ subsector \cite{Beisert:2003ys}.
 
With higher loop integrability, one can look for a set of Bethe equations.
This was done first by Serban and Staudacher, where they argued that the
Inozemtsev chain was consistent with the known dilatation operator up
to three-loops in the $SU(2)$ sector \cite{Serban:2004jf}.  
With these modified Bethe equations, Serban and 
Staudacher were able to compute the anomalous dimensions for the coherent
operators
dual to the circular and spinning string in the $SU(2)$ sector and showed that
 it agreed with the predicted semiclassical string values at
two loops but failed to match at three loops \cite{Serban:2004jf}.  It
was subsequently shown that all coherent $SU(2)$ solutions match the string
predictions at two-loops \cite{Kazakov:2004qf}.  It was also shown that
the effective action of the spin chain matches the sigma model action
at two loops \cite{Kruczenski:2004kw}.   Another proposed
set of Bethe equations has also been put forth, which further assumes
that BMN scaling \cite{Berenstein:2002jq} 
is present at all orders in perturbation theory \cite{Beisert:2004hm}.  
This matches the Inozemtsev chain to three loop order
but diverges from it at 4 loops.

It is also possible to make one-loop comparisons outside of the $SU(2)$
sector.  This was first done in \cite{Beisert:2003xu}, where it was shown that
a solution of the full $SO(6)$ Bethe equations was consistent with
the spectrum of a semiclassical pulsating string \cite{Minahan:2002rc}. This
analysis was carried out for more general scenarios in 
\cite{Engquist:2003rn,Kristjansen:2004ei},
where a continuous set of $SO(6)$ and $SU(3)$ solutions were found and
shown to agree with the string predictions at one-loop
\cite{Frolov:2003qc,Arutyunov:2003uj,Engquist:2003rn,Arutyunov:2003za}. It was also
recently shown how to compare the effective action for the one-loop
$SU(3)$ chain with the string model, where agreement was found, essentially
confirming that all coherent one-loop solutions will match with the
string predictions \cite{Hernandez:2004uw,Stefanski:2004cw}.

A natural question to ask is how the analysis for the
coherenet operators in the full $SO(6)$ 
or $SU(3)$ sectors can be extended to higher loops.  At first glance
this would seem problematic, since these sectors are  not closed above
one loop.  For example, in the $SU(3)$ sector three different scalar fields
can mix into two fermion fields, preserving spin, $R$-charge and the bare
dimension \cite{Beisert:2003ys}.    
In the $SO(6)$ sector, the problem is even more acute;
there can be mixing into the full $SU(2,2|4)$ set of fields.

However, in the semiclassical limit 
this mixing should be suppressed.  The
states in the semiclassical limit are coherent; the quantum states at
the individual spin sites vary slowly  over
the length of the chain.  The mixing outside of the $SU(3)$ or $SO(6)$ is
essentially a quantum fluctuation that is suppressed by $1/L$.
In the case of  coherent $SU(3)$ states, we will explicitly show this.   
In the case of $SO(6)$
in the semiclassical limit, it is possible to find solutions that are
further restricted to an $SO(4)$ subgroup, which is normally not closed even
at one loop.  Again, however, one expects mixing outside of this subgroup
to be suppressed by $1/L$.

If the mixing is suppressed, then it is natural to assume that the
$SO(6)$ or $SU(3)$ Bethe equations are modified in a way analagous
to the $SU(2)$ equations.  For the $SO(6)$ solutions, we will show
for a general class of solutions  that the integral 
equations reduce to two independent $SU(2)$ integral equations.  Each
$SU(2)$ equation is then modified in the way proposed by Serban
and Staudacher \cite{Serban:2004jf}.  
A similar modification is done for the $SU(3)$ equations.
In these cases
 we will show that the anomalous dimensions are consisistent with
the dual string predictions up to two loop order but in general fail
at three loops.  In both cases there is one exceptional point where
the three-loop predictions agree.

The three-loop failure is to be expected, based on past failures
of $1/L$ corrections to the BMN limit 
\cite{Callan:2003xr,Callan:2004uv,Callan:2004ev}, or for other semiclassical 
predictions \cite{Beisert:2003ea,Serban:2004jf,Beisert:2004hm}.  
However, in our case we have an adjustable parameter
and we find that there is one value, aside from the BMN limit, where there is
three-loop agreement.

The paper is organized as follows.  In section 2
we show why coherent $SU(3)$ operators have their
 mixing into the fermion sector suppressed by a factor of
$1/L$.
In section 3  we show how a class of $SO(6)$ solutions reduce to
$SO(4)$ solutions.  In section 4 we consider rational examples
for these reduced $SO(6)$  solutions, 
which includes duals to pulsating strings with an $R$-charge $J$.
  In section 5, we consider higher loop terms for the anomalous
dimension, showing that the two loop prediction
for pulsating string duals matches the semiclassical
string prediction. We then show that the two loop $SU(3)$ solution with
$R$-charge assignment $(J',J',J)$
is consistent with the string prediction.  Next we compute the three loop
terms for both cases, finding in general that the string and gauge
theory predictions do not agree, except when $J=L/3$, where $L$ is
the bare dimension.  In section 6 we compute the fluctuation spectra
of the pulsating strings, using the fact that they are essentially
reduced to the $SO(4)$ sector.
  In section 7 we present our conclusions.  An appendix contains
some more complicated expressions.

\sectiono{Mixing suppression for coherent $SU(3)$ states.}

In this section we demonstrate that coherent operators made up of three chiral 
scalar
fields have suppressed mixing to operators 
with fermions. 
As was already stated, the $SU(3)$ sector is not closed under mixing,
but is instead enlarged to  the sector $SU(2|3)$, where the mixing can first 
appear at the
two-loop level \cite{Beisert:2003ys}.  To see this, we note that there
is a two loop process where three scalar fields are converted to
two fermion fields, as is shown in figure 1(a).  There is also
a one-loop process where two fermions convert to 3 scalars, as
is shown in figure 1(b). Both processes come with a factor of $g^3$,
but they also come with a factor of $N^{3/2}$.  This is because
the two loop term has a factor of $N^2$ but this is divided by a
factor of $N^{1/2}$ since the operator with the extra three scalars has
one more field than the operator with two fermions and each field comes
with a normalization factor of $N^{-1/2}$.  The overall effect of
these mixing terms is a shift of the anomalous dimension by a term
of  $\lambda^2$.

\begin{figure}[tp]
\centerline{\epsfig{file= 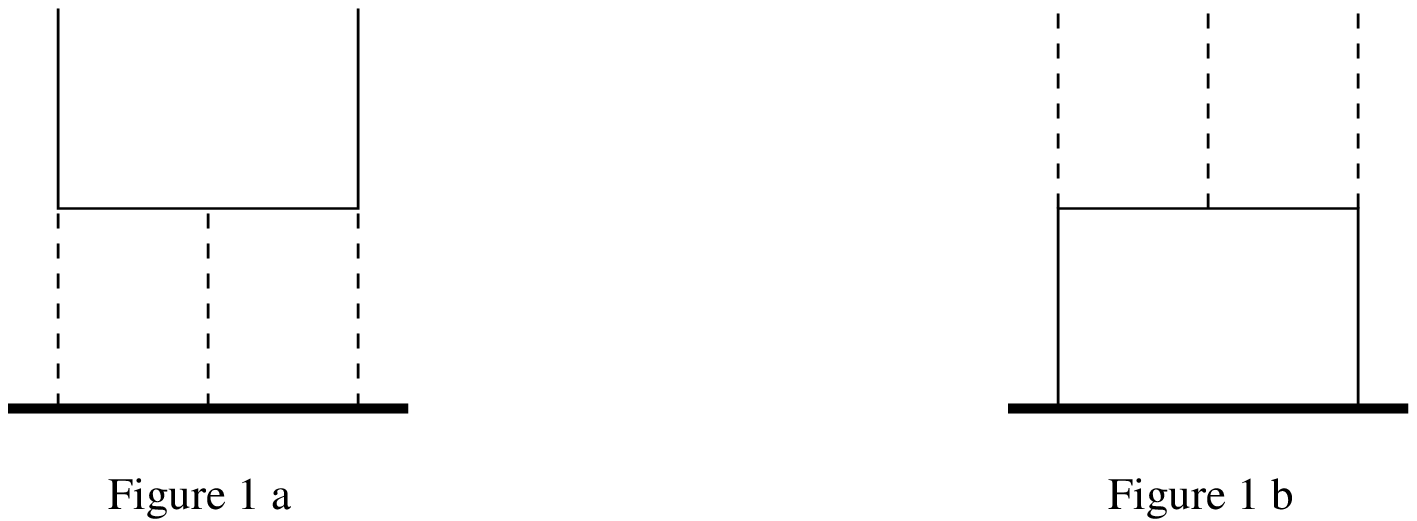,width=150mm}}
\caption{(a) Mixing of three scalars (dashed lines) into two fermions (solid
lines).  (b) Mixing of two fermions into three scalars}
\label{mixing}
\end{figure}

Under the $SU(3)\times U(1)$ subgroup of $SU(4)$, the 
chiral scalars transform 
in the ${\bf 3_{+1}}$ representation, while the fermions  
in the ${\bf(2,1)}$ representation of the Lorentz group transform in
the ${\bf 1_{3/2}}+{\bf 3_{-1/2}}$ representation.   Hence the antisymmetrized
triplet of scalars  
can mix
with the Lorentz singlet of two $SU(3)$ singlet fermions.
The contribution to the mixing matrix is a sum of local terms of the form
\cite{Beisert:2003ys}
\begin{equation}\label{3to2}
H_3=C\vareps_{\alpha\beta}\vareps^{abc}\left\{\alpha\beta\atop abc\right\}
+C^*
\vareps_{abc}\vareps^{\alpha\beta}\left\{ abc\atop\alpha\beta\right\}
\end{equation}
where $\alpha$ and $\beta$ are the $SU(2)$ Lorentz indices and $a,b,c$ 
are the $SU(3)$ flavor indices. $C$ is constant.
The bottom line in the brackets refers to  three (two) neigboring fields
in the {\it in} state and the top line refers to  two (three)
neighboring fields in the {\it out} state. The effect of this term is
to make the chain dynamical since it either increases or decreases
the number of sites in the chain by one \cite{Beisert:2003ys}.

Now suppose we consider the $SU(3)$ chain with nearest neighbor interactions
in the classical limit.  This
was recently described in \cite{Hernandez:2004uw,Stefanski:2004cw}.  
The idea is to write the states
in terms of collective coordinates.  For an $SU(3)$ transformation
on a state in the
fundamental representation, there is an $SU(2)$ subgroup that leaves the
state invariant, plus a $U(1)$ that multiplies the state by a phase.
Hence the collective coordinates are coordinates on the coset 
$SU(3)/SU(2)\times U(1)$.  These are described by 4 angles, with  a general
state   written as \cite{Hernandez:2004uw,Stefanski:2004cw}
\begin{equation}
|\vec n(\s)\rangle=\cos\theta(\s)\cos\psi(\s) e^{i\varphi(\s)}|1\rangle
+\cos\theta(\s)\sin\psi(\s) e^{-i\varphi(\s)}|2\rangle
+\sin\theta(\s) e^{i\phi(\s)}|3\rangle,
\end{equation}
where $\phi$ and $\varphi$ range between $0$ and $2\pi$ and
$\theta$ and $\psi$ range between $0$ and $\pi/2$.  $\s$ labels the
site on the chain, with neighboring sites differing by $\Delta\s=2\pi/L$.
The action can then be determined by considering the inner product
$\langle\vec n(\s+\Delta\s)|\vec n(\s)\rangle$, 
from which one can determine the equations
of motion \cite{Hernandez:2004uw,Stefanski:2004cw}

Consider then the action of $H_3$ in 
(\ref{3to2}) on the state $|\vec n(\s-\Delta\s)\rangle|\vec n(\s)\rangle|\vec n(\s+\Delta\s)\rangle$
with the angles slowly varying between the sites.  
The resulting state has the form
\begin{equation}
H_3|\vec n(\s-\Delta\s)\rangle|\vec n(\s)\rangle|\vec n(\s+\Delta\s)\rangle
=C(\Delta\s)^3e^{i\phi(x)}
F(\theta(x),\psi(x),\varphi(x),\phi(x))\vareps_{\alpha\beta}|\alpha\rangle|\beta\rangle,
\end{equation}
where $F(\theta(x),\psi(x),\varphi(x),\phi(x))$ is a sum of three
derivative terms and 
$|\alpha\rangle$ and $|\beta\rangle$ are the fermion states at
neighboring sites.
 In the thermodynamic limit $F$ is finite for $H_3$ acting on a coherent
state.

At the one-loop level, an operator with chiral scalars and two fermion fields
is closed; it only mixes with operators with
two fermion fields.  In fact the state with two fermions can be thought
of as a fluctuation from a state with no fermions.  There are of order
$L$ such fluctuations corresponding to the different choices of
momentum for the fermions.  These fluctuations  increase the 
anomalous dimension by 
order $\lambda n^2/L^2$, with $n$ an integer ranging from 1 to order $L$.
The state with the fermions next to each other on the chain is approximately
a linear combination of the momentum states with each coefficient
$1/L$.  But there are $L$ sites on the chain for the transition to happen
so the matrix element between a zero fermion state and a two fermion state
is approximately
\begin{equation}
(gN)^{3/2}\langle 2|H_3|0\rangle \sim \left(\frac{\lambda}{L^2}\right)^{3/2}
\end{equation}
Hence we expect these fermion fluctuations to change the anomalous dimension
by order
\begin{equation}
\sum_n \frac{(\lambda/L^2)^{3}}{n^2\lambda/L^2}\sim \frac{\lambda^2}{L^4}.
\end{equation}
In the semiclassical limit, the two-loop contribution is of order
$\lambda^2/L^3$, so the contribution from the fermion fluctuations is
suppressed by a factor of $1/L$.

\sectiono{$SO(6)$ reduction to $SO(4)$}

In this section we show how a certain class of solutions to the
$SO(6)$ Bethe equations reduce to $SO(4)\simeq SU(2)\times SU(2)$ 
Bethe equations. The arguments appearing here are essentially a generalization
of those in \cite{Engquist:2003rn}.  
The $SO(4)$ symmetry can be easily understood
from the string side, where it has been shown that the semiclassical
string duals are restricted to an $R\times S_3$ subspace 
\cite{Minahan:2002rc,Engquist:2003rn,Kazakov:2004qf}.  The
$SO(4)$ symmetry corresponds to the isometry group of $S_3$.

We will consider single trace operators $\OO$
 made up of scalar
fields only.  The operators are not holomorphic; they  
 contain $\overline{X}$, $\overline{Y}$ and $\overline{Z}$ scalar
fields inside the trace.   We will assume that the operators are highest
weights of  $SO(6)$ representations with bare dimension $L$ and
with $R$-charges
$(0,0,J)$.  In terms of $SO(6)$ Dynkin indices, these 
representations are denoted by $[0,J,0]$.

The $SO(6)$ Bethe equations are given by
\begin{eqnarray}\label{betheSO6}
\left(\frac{u_{1,i}+i/2}{u_{1,i}-i/2}\right)^L&=&
\prod_{j\ne i}^{n_1}\frac{u_{1,i}-u_{1,j}+i}{u_{1,i}-u_{1,j}-i}
\prod_{j}^{n_2}\frac{u_{1,i}-u_{2,j}-i/2}{u_{1,i}-u_{2,j}+i/2} 
\prod_{j}^{n_3}\frac{u_{1,i}-u_{3,j}-i/2}{u_{1,i}-u_{3,j}+i/2}
\nonumber\\
1&=&\prod_{j\ne i}^{n_{2}}\frac{u_{2,i}-u_{2,j}+i}{u_{2,i}-u_{2,j}-i}
\prod_{j}^{n_{1}}\frac{u_{2,i}-u_{1,j}-i/2}{u_{2,i}-u_{1,j}+i/2}
\nonumber\\
1&=&\prod_{j\ne i}^{n_{3}}\frac{u_{3,i}-u_{3,j}+i}{u_{3,i}-u_{3,j}-i}
\prod_{j}^{n_{1}}\frac{u_{3,i}-u_{1,j}-i/2}{u_{3,i}-u_{1,j}+i/2}\,,
\end{eqnarray}
 where  $n_1$, $n_2$ and $n_3$ denote the number of Bethe roots associated
with each simple root of $SO(6)$.  For this choice, the Dynkin indices
of this representation
are given by \mbox{$[n_1-2n_2, L-2n_1+n_2+n_3,n_1-2n_3]$}.  
The anomalous dimension is only directly related to the $u_1$ roots 
and is given by
\be\label{anomdim}
\gamma\,=\,\frac{\lambda}{8\pi^2}\sum_{j=1}^{n_1}\frac{1}{(u_{1,j})^2+1/4}.
\ee
There is also a momentum condition,
\begin{equation}\label{momc}
1\,=\,\prod_{j=1}^{n_1}\frac{u_{1,j}+i/2}{u_{1,j}-i/2}\,,
\end{equation}
which is a consequence of the
cyclicity of the trace in $\OO$.
We now consider the thermodynamic limit, where  $n_i\sim L$.  We
also assume a ``half-filling'' condition for the roots, where $n_2=n_3=n_1/2$
and we further assume that the distribution of the $u_2$ roots is the 
same as the distribution of the $u_3$ roots, which clearly is consistent
with (\ref{betheSO6}). 

As was argued in \cite{Beisert:2003xu,Engquist:2003rn,Kazakov:2004qf}, 
the roots will lie along cuts in the complex
plane.  We now assume that the $u_1$ roots are on multiple cuts,
but the $u_2$ and $u_3$ roots are on a single cut.
Taking logs on both sides of (\ref{betheSO6}) and rescaling by $u=xL$,
the Bethe equations reduce to the integral equations
\begin{eqnarray}\label{inteqs}
\frac{1}{x}-2\pi n_i&=& 2\pint_{\CC_i}dx'\frac{\s(x')}{x-x'}+2\sum_{j\ne i}\int_{\CC_j}dx'\frac{\s(x')}{x-x'}
- 2\int_{\CC'}dx'\frac{\rho(x')}{x-x'}\qquad x\in \CC_i\nonumber\\
0&=& 2\pint_{\CC'}dx'\frac{\rho(x')}{x-x'}-\sum_j\int_{\CC_j}dx'\frac{\s(x')}{x-x'}\qquad x\in \CC',\end{eqnarray}
where $\CC_i$ are the cuts for the rescaled $u_1$ roots, $\CC'$ is the cut for 
the rescaled $u_2$ and  $u_3$ roots, and $n_i$ labels the log branch.  Roots
along the same cut are on the same log branch.  The root densities
satisfy normalization conditions
\be\label{rootdens}
\sum_j\int_{\CC_j}\s(x')dx'=2\int_{\CC'}\rho(x')dx'=\frac{n_1}{L}=\frac{L-J}{L}
\equiv\alpha\,,
\ee
and the anomalous dimension and momentum condition in (\ref{anomdim}) and
(\ref{momc}) become
\begin{equation}\label{anomdimint}
\gamma=\frac{\lambda}{8\pi^2}\sum_{j}\int_{\CC_j}\frac{dx\s(x)}{x^2}\,,
\end{equation}
and
\begin{equation}\label{momcint}
2\pi m\,=\,\sum_j\int_{\CC_j}\frac{dx\s(x)}{x}\,.
\end{equation}

With the conditions in (\ref{rootdens}), we can Hilbert transform
the second equation in (\ref{inteqs}) to get the relation for $\rho(x)$
\be\label{rhoeq}
\rho(x)=-\frac{1}{2\pi^2}\sqrt{(x-a)(x-b)}
\pint_{a}^{b}\frac{dx'}{x-x'} \frac{1}{\sqrt{(x'-a)(x'-b)}}\sum_j\int_{\CC_j}dx''\frac{\s(x'')}{x'-x''}
\ee
where $a$ and $b$ are the end points of the cut $\CC'$.  Since the existence of
a Bethe root requires the presence of its complex conjugate, we must have
$b=a^*$. If we invert the
order of integration and integrate over $x'$ by deforming the contour, we find
 \be\label{rho2eq}
\rho(x)=-\frac{1}{2\pi i}\sum_j\int_{\CC_j} dx''\frac{\s(x'')}{x-x''}
\frac{\sqrt{(x-a)(x-a^*)}}{\sqrt{(x''-a)(x''-a^*)}}.
\ee
If we now reinsert $\rho(x)$ in (\ref{rhoeq}) into (\ref{rootdens}), then
\begin{eqnarray}\label{rhoeq2}
\int_{\CC'}dx'\rho(x')&=&\frac{1}{2}\sum_j\int_{\CC_j}dx'\s(x')
+\frac{1}{2}\sum_j\int_{\CC_j}\frac{dx\s(x')(a+a^*-x')}{\sqrt{(x'-a)(x'-a^*)}}
=\frac{\alpha}{2}
\end{eqnarray}
Comparing (\ref{rhoeq2}) with (\ref{rootdens}) we see that $a\to\infty$ 
while at the same time, ${\rm Re}\,a/{\rm Im}\,a\to0$ and so
$\CC'$ essentially cuts the complex plane in two.  In this case
$\frac{\sqrt{(x-a)(x-a^*)}}{\sqrt{(x'-a)(x'-a^*)}}\to\eps(x,x')$ where
$\eps(x,x')=\pm 1$ with the $+$ ($-$) sign if $x$ and $x'$ are on the
same (opposite) sides of $\CC'$.  If we now take this limit 
on eq. (\ref{rho2eq}), we obtain
\begin{equation}
\int_{\CC'}\frac{dx'\rho(x')}{x-x'}=\sum_{j'}\int_{\CC_{j'}}\frac{dx'\s(x')}{x-x'},
\end{equation}
where the sum over the index $j'$ refers to those cuts that are on the
{\it opposite} side of $\CC'$ from $x$.  

If we now examine (\ref{inteqs}),
we see that the effect of the roots on $\CC'$ is to screen the cuts
on either side from each other.  In other words, the system has degenerated
 to two independent sets of roots, each satisfying $SU(2)$ spin $1/2$ 
thermodynamic Bethe equations.  The contribution to the anomalous dimension
is the sum of the contribution from each $SU(2)$ sector.   The same is
true for the momentum.  In fact, we now see that the momentum condition 
in (\ref{momcint}) does
not apply to each $SU(2)$ individually, but only to their combination.
These arguments may also be applied to operators containing fields 
transforming in $SU(2,2)$ representations, where in the thermodynamic
limit the $SU(2,2)$ can degenerate into $SU(1,1)\times SU(1,1)$ 
\cite{Smedback:2004yn}.
 
As with the case for a single $SU(2)$, it is convenient to consider
the resolvants
\begin{eqnarray}
G_\pm(x)&=&\sum_{j_{\pm}}\int_{\CC_{j_{\pm}}}\frac{dx'\s_\pm(x')}{x-x'}
\end{eqnarray}
where the $+$ ($-$) refers to the roots on the right (left) of $\CC'$.
The resolvents satisfy the equations
\begin{equation}
G_\pm(x+i0)+G_\pm(x-i0)=\frac{1}{x}-2\pi n_{j_\pm}\qquad\qquad x\in \CC_{j_\pm}
\end{equation}
where the $n_{j_-}<n_{j_+}$ for all $j_-$ and $j_+$.  
In order to insure the 
cyclicity of the trace in $\OO$, the resolvents must satisfy
\begin{equation}\label{trace}
G_+(0)+G_-(0)=-2\pi m
\end{equation}
where $m$ is an integer.
Likewise, from (\ref{anomdim}) the anomalous dimension is
given by
\begin{eqnarray}\label{anomdim2}
\gamma=-\frac{\lambda}{8\pi^2L}\left(G_+'(0)+G_-'(0)\right).
\end{eqnarray}
Finally, equation (\ref{rootdens}) leads to the asymptotic condition
\begin{equation}\label{Gasymp}
G_+(x)+G_-(x)\approx \frac{\alpha}{x},\qquad x\to\infty.
\end{equation} 

\sectiono{Rational examples}

The simplest situation to consider is when there is a single cut
contributing to each resolvent.  In this case $G_+(x)$ and $G_-(x)$
have an algebraic solution.  Let us assume that 
\begin{equation}
-G_\pm(0)=2\pi s_\pm
\end{equation} 
and that the cuts are on the branches $n_+$ and $n_-$.
In order to satisfy the trace condition in (\ref{trace}), we have
$s_++s_-=m$.

 The resolvents are now given by \cite{Kazakov:2004qf}
\begin{equation}\label{Grat}
G_\pm(x)=\frac{1}{2x}\left(1+\sqrt{(2\pi n_\pm x)^2+4\pi(2s_\pm-n_\pm)+1}\right)
-\pi n_\pm
\end{equation}
where the branch of the square root is chosen to cancel the pole at $x=0$.
The asymptotic behavior for $G_\pm(x)$ is
\begin{equation}
G_\pm(x)\approx\frac{s_\pm/n_\pm }{x},
\end{equation}
and so comparing with (\ref{Gasymp}) we see that $\alpha=(s_+/n_+)+(s_-/n_-)$.
Thus, in terms of $\alpha$ and $m$ we have
\begin{equation}
s_\pm=-\frac{(\alpha n_\mp-m)n_\pm}{n_\pm-n_\mp}
\end{equation}
Note that $s_\pm/n_\pm>0$ in order for the states to be physical, but unlike
the case of one $SU(2)$, it is possible to have $s_+/n_+>1/2$ or $s_-/n_->1/2$,
as long as $\alpha\le 1$.

Using (\ref{anomdim2}), we see that the anomalous dimension is
\begin{equation}
\gamma=\frac{\lambda}{2L}\frac{2\alpha(1-\alpha)n_+^2n_-^2-(\alpha n_+n_-+m^2)(n_+^2+n_-^2)+m(n_+^3+n_-^3-(1-2\alpha)n_+n_-(n_++n_-))}{(n_+-n_-)^2}.
\end{equation}
This is a more general solution than that given in \cite{Engquist:2003rn}, 
where the solutions
there correspond to $n_+=-n_-$, $m=0$.  On the string side, these
solutions were recently described in \cite{Arutyunov:2003za,Dimov:2004xi}.

An interesting application would be to compute the $1/L$ corrections
along the lines of \cite{Lubcke:2004dg}.

\sectiono{Higher loops}

\subsection{$SO(6)$ rationals at two loops}

The two-loop dilatation operator has yet to be determined in the full $SO(6)$
sector.  However, it is known in the $SU(2)$ subsector, so it is natural
to just use the $SU(2)$ results, assuming again that in the semiclassical
limit an $SO(6)$ spin chain state can reduce to an $SO(4)$ spin chain
solution.

At two loops the $SO(6)$
sector is not closed under dilatations.  But as per our discussion
in section 2, we will assume that these effects
are subdominant in the semiclassical limit, and that the mixing outside
the $SO(6)$ sector is suppressed by factors of $1/L$.

The two-loop modification for an $SU(2)$ chain leads to the modified
integral equations for the resolvents \cite{Serban:2004jf,Kazakov:2004qf}
\begin{equation}
\label{BAETL}
{1\over x} +{2T\over x^3}   -2\pi n_j=
G_\pm(x+i0)+ G_\pm(x-i0)\qquad x\in \CC_{j_\pm}.
\end{equation}
where $T=\frac{\lambda}{16\pi^2 L^2}$. 
The momentum to two loop order is given by
\begin{equation}
2\pi s_\pm=-G_\pm'(0)-TG_\pm''(0)
\end{equation}
and the anomalous dimension is
\begin{equation}
\gamma=-2TL\left(G_+'(0)+\frac{T}{2}G_+'''(0)+G_-'(0)+\frac{T}{2}G_-'''(0)\right)
\end{equation}

In \cite{Kazakov:2004qf} the two loop result for rational solutions was 
explicitly computed.
Borrowing those results, we find that the two-loop contribution to $\gamma$
for the
rational $SO(6)$ solution of the last section is
\begin{equation}
\label{gamma2}
\gamma_2=-\frac{\lambda^2}{8L^3}\left[s_+(n_+-s_+)(n_+^2-3s_+(n_+-s_+))
+s_-(n_--s_-)(n_-^2-3s_-(n_--s_-))\right].
\end{equation}
In the case where $n_\pm=\pm n$, $s_\pm =\pm n\alpha/2$, (\ref{gamma2})
simplifies to
\begin{equation}
\label{gamma2pulse}
\gamma_2=-\frac{\lambda^2m^4}{64L^3}\alpha(2-\alpha)(4-3\alpha(2-\alpha))
\end{equation}

We now claim that the result in (\ref{gamma2pulse}) is consistent
with the semiclassical solution of a string pulsating on $S_5$ with
its center of mass revolving around an equator with angular
momentum $J$.  The discussion follows closely that of 
\cite{Engquist:2003rn} and previous work in \cite{Minahan:2002rc}.
Let us consider a circular pulsating string expanding
and contracting on $S_5$,  and with a center of mass that
is moving on an $S_3$ subspace.  We will assume that the string is
fixed on the spatial coordinates in $AdS_5$, so the relevant metric is
\be\label{S5met}
ds^2\ =\ R^2(-dt^2+\sin^2\theta\ d\psi^2+\ d\theta^2\ +\ \cos^2\theta\ d\Omega_3^2),
\ee
where $d\Omega_3$ is the metric on the $S_3$ subspace and 
$R^2=2\pi\alpha'\sqrt{\lambda}$.
The string is stretched only along the $\psi$ coordinate and is wrapped
$n$ times.  
Fixing a gauge $t=\tau$, the Nambu-Goto action  
reduces to
\be\label{NGsph}
S\ =-\ n\sqrt{\la}\int dt\ \sin\theta\ \sqrt{1-\dot\theta^2-\cos^2\theta g_{ij}
\dot\phi^i\dot\phi^j},
\ee
where $g_{ij}$ is the metric on $S_3$ and $\phi^i$ refers to the coordinates
on $S_3$.
In \cite{} it was shown that the Hamiltonian is
\be
H\ =\ \sqrt{\Pi_\theta^2+\frac{g^{ij}\Pi_i\Pi_j}{\cos^2\theta}+n^2\la\sin^2\theta},
\ee
where  $\Pi_\theta$ and $\Pi_i$ are the canonical momentum for
$\theta$ and the angles on $S_3$. 
The square of $H$ has the form of  a Hamiltonian for a particle on $S_5$
with an angular dependent potential $n^2\la\sin^2\theta$.  
The semiclassical limit corresponds to large quantum numbers for the
canonical momenta, so the potential may be considered as a perturbation.
This string configuration has a total $S_5$ angular momentum $L$,
which is the bare dimension for its gauge dual operator.  On an $S_3$ subspace
the angular momentum is $J$, which is the $R$-charge for the dual
operator.

The unperturbed wave-functions are solutions to the Schroedinger equation
\be
E^2\Psi(w)=-\frac{4}{w}\frac{d}{dw}w^2(1-w)\frac{d}{dw}\Psi(w)+
\frac{J(J+2)}{w}\Psi(w),
\ee
where $w=\cos^2\theta$.
We will assume that $J$ and $L$ are even
and define $j=J/2$ and $\ell=L/2$.
The normalized $S_5$ wave functions are then given by
\be
\Psi_{\ell,j}(w)=\frac{\sqrt{2(\ell+1)}}{(\ell-j)!}\frac{1}{w^{j+1}}\left(\frac{d}{dw}\right)^{\ell-j}w^{\ell+j}(1-w)^{\ell-j}.
\ee
In \cite{} the first order correction to $E^2$ was shown to be
\be\label{1storder}
\int_0^1wdw\Psi_{\ell,j}(w)n^2\lambda(1-w)\Psi_{\ell,j}(w)= n^2\lambda 
\frac{2(\ell+1)^2-(j+1)^2-j^2}{(2\ell+1)(2\ell+3)}
\approx\frac{n^2\lambda(\ell^2-j^2)}{2\ell^2}.
\ee
For the second order correction, the  matrix elements 
$\langle\Psi_{\ell',j'}|(1-w)|\Psi_{\ell,j}\rangle$,  
$\{\ell',j'\}\ne\{\ell,j\}$
 satisfy
\begin{eqnarray}\label{2ndorder}
\langle\Psi_{\ell',j'}|(1-w)|\Psi_{\ell,j}\rangle
&=&\int_0^1wdw\Psi_{\ell',j}(w)(1-w)\Psi_{\ell,j}(w)\,\delta_{j',j}
\nonumber\\
&=&\frac{1}{2}\left(\frac{(\ell'+j+1)(\ell'-j)}{(2\ell'+1)\sqrt{\ell'(\ell'+1)}}\delta_{\ell',\ell+1}
+\frac{(\ell+j+1)(\ell-j)}{(2\ell+1)\sqrt{\ell(\ell+1)}}\delta_{\ell'+1,\ell}\right)\delta_{j',j}
\nonumber\\
&\approx&\left(\frac{\ell^2-j^2-j}{4\ell^2}
(\delta_{\ell',\ell+1}+\delta_{\ell'+1,\ell})+
\frac{j^2}{4\ell^3}\left(3\delta_{\ell',\ell+1}+\delta_{\ell'+1,\ell}\right)
\right)\delta_{j',j}
\end{eqnarray}
Thus, up to second order in $\lambda/L^2$ and
assuming large $L$, $J$ and $\lambda$, $E^2$ is given by
\be
E^2\ =\ L^2\  +\ n^2\lambda\frac{L^2-J^2}{2L^2}\ +\ (n^2\lambda)^2
\frac{(L^2-J^2)(L^2-5J^2)}{32L^6}\ + O(\lambda^3/L^4)
\ee
Hence, again up to second order in $\lambda/L^2$, $E$ is
\be\label{Estring}
E=L\ +\ n^2\lambda\frac{L^2-J^2}{4L^3}\ -\ (n^2\lambda)^2
\frac{(L^2-J^2)(L^2+3J^2)}{64L^7}\ + O(\lambda^3/L^5)
\ee
Replacing $J=L(1-\alpha)$ and 
comparing (\ref{Estring}) to (\ref{gamma2pulse}), we find agreement.

\subsection{The $SU(3)$ chain at two loops}

Operators $\OO$ containing the three complex scalar fields but not their
conjugates have a one-loop dilatation operator that maps to a Hamiltonian
for an integrable $SU(3)$ chain \cite{Minahan:2002ve}.
The dilatation operator is
known to three loops and has been shown to be consistent with 
integrability \cite{Beisert:2003ys}.  
However, the $SU(3)$ operators are not closed
at higher loops, instead they mix under an $SU(2|3)$ subgroup of
$SU(2,2|4)$.

However, based on our earlier arguments we will  assume that 
 mixing with operators
containing fermion fields can be ignored in the semiclassical limit.  We will only consider states with 
$R$-charge
assignment $(J',J',J)$ and bare dimension $L=J+2J'$.  This corresponds
to having no $u_3$ roots and half as many $u_2$ roots as $u_1$ roots 
\cite{Engquist:2003rn}.
We then make the ansatz that the two loop modification for the Bethe
equations  changes the lhs of the first equation 
in (\ref{inteqs}) to the  lhs of the modified
$SU(2)$ Bethe equations in \cite{Serban:2004jf}.  The upshot of all this
is that the first equation in (\ref{inteqs}) gets modified to
\begin{equation}
\frac{1}{x}+\frac{2T}{x^3}-2\pi n_i= 2\pint_{\CC_i}dx'\frac{\s(x')}{x-x'}+2\sum_{j\ne i}\int_{\CC_j}dx'\frac{\s(x')}{x-x'}
- \int_{\CC'}dx'\frac{\rho(x')}{x-x'}\qquad x\in \CC_i.
\end{equation}
Let us suppose that the $u_1$ roots are distributed symmetrically
on two cuts, $\CC_+$ and $\CC_-$,
 on either side of $\CC'$ which lies on the imaginary axis.  The branch
numbers 
are assumed to be $n_+=-n_-=n$.  Then the resolvent $W(x)$,
\begin{equation}
W(x)=\int_{\CC_+}\frac{dx'\s(x')}{x-x'},
\end{equation}
satifies the Riemann-Hilbert equation \cite{Engquist:2003rn}
\begin{equation}\label{Weq}
W(x+i0)+W(x-i0)-W(-x)=\frac{1}{x}+\frac{2T}{x^3}-2\pi n\qquad x\in\CC_+.
\end{equation}
As in \cite{Engquist:2003rn}, we write $W(x)=W_r(x)+w(x)$, where
\begin{equation}
W_r(x)=\frac{1}{3x}+\frac{2T}{3x^3}-2\pi n,
\end{equation}
and $w(x)$ satisfies the homogeneous Riemann-Hilbert equation
\begin{equation}\label{whom}
w(x+i0)+w(x-i0)-w(-x)=0\qquad x\in \CC_+.
\end{equation}
Then the function
\begin{equation}\label{req}
r(x)=w^2(x)-w(x)w(-x)+w(-x)^2
\end{equation}
is even and regular across $\CC_+$.  In terms of the filling fraction
$\alpha=2J'/L$, $w(x)$ is asymptotically
\begin{eqnarray}\label{wasymp}
w(x)&\approx& (2\pi n)+\left(\frac{\alpha}{2}-\frac{1}{3}\right)\frac{1}{x}\qquad
x\to\infty
\nonumber\\
w(x)&\approx&-\frac{1}{3x}-\frac{2T}{x^3}\qquad\quad x\to0.
\end{eqnarray}
Then, in order to be consistent with (\ref{wasymp}),
 $r(x)$ is given by
\begin{equation}
r(x)=(2\pi n)^2+\frac{1}{3x^2}+\frac{4T}{3x^4}+\frac{pT}{x^2},
\end{equation}
to linear order in $T$, where the coefficient $p$ is to be determined.

One then has the equation
\begin{equation}
w^3(x)-r(x)w(x)=w^3(-x)-r(-x)w(-x)\equiv s(x)
\end{equation}
where $s(x)$ is an odd function that is also regular across the cut
since $w(-x)$ is regular there.  The asymptotic conditions for $w(x)$
then give
\begin{equation}
s(x)=\frac{2}{27x^3}+(2\pi n)^2\left(\alpha-\frac{2}{3}\right)\frac{1}{x}
+\frac{4T}{9x^5}+\frac{qT}{x^3}
\end{equation}
to linear order in $T$, where $q$ is to be determined.  

It is convenient to use $g(x)=w(x)-w(-x)$, in which case $g(x)$ satisfies
the equation
\begin{equation}\label{gxeq}
g^3(x)-r(x)g(x)+s(x)=0.
\end{equation}
and the full resolvant $G(x)$ is given by $G(x)=g(x)+W_r(x)-W_r(-x)$.
Solving (\ref{gxeq}) as an expansion about $x=0$, we find that up to
linear order in $T$, $g(x)$ is given by
\begin{equation}
g(x)=-\frac{4T}{x^3}-\frac{2}{3x}-\alpha(2\pi n)^2\left(1-4T(1-\alpha)(2\pi n)^2+pT\right)x-\alpha(1-2\alpha)(2\pi n)^4+...
\end{equation}
If we substitute the corresponding expression for $w(x)$ into
(\ref{req}), we find that \mbox{$p=2\alpha(2\pi n)^2$} to leading order in $T$.
It then follows that $q=\frac{4}{3}(2\pi n)^2(2\alpha-1)$.

The anomalous dimension to two loop order is now found to be
\begin{equation}
\gamma=-2TL\left(G'(0)+\frac{T}{2}G'''(0)\right)=\frac{n^2\alpha\lambda }{2L}\left(1-
\frac{n^2\lambda}{4L^2}\right).
\end{equation}
This agrees with the result of Frolov and Tseytlin in \cite{Frolov:2003qc}.

\subsection{Three loops}

In this subsection we consider the 3-loop contributions for the pulsating
string and the $SU(3)$ circular string.  More details can be found in
the appendix.  As with the $SU(2)$ examples,  the gauge predictions
are in general different from the string predictions.  However, both
types of operators have a continuous parameter that can be adjusted,
and curiously there is agreement at one point in each example. 

At three loops, using the $SU(2)$ results of the
Inozemtsev chain \cite{Serban:2004jf}, the integral equations are modified to
\begin{equation}
\frac{1}{x}+\frac{2T}{x^3}+\frac{6T^2}{x^5}-2\pi n_i= 2\pint_{\CC_i}dx'\frac{\s(x')}{x-x'}+2\sum_{j\ne i}\int_{\CC_j}dx'\frac{\s(x')}{x-x'}
-\nu \int_{\CC'}dx'\frac{\rho(x')}{x-x'}\qquad x\in \CC_i.
\end{equation}
where $\nu=2$ for the $SO(6)$ chain and $\nu=1$ for the $SU(3)$ chain.
The modifications to the momentum and the anomalous dimension to third
order in $\lambda/L^6$ are
\begin{eqnarray}\label{3gamma}
2\pi s&=& -G(0)-G''(0)T-\frac{1}{4}G^{(4)}(0)T^2\nonumber\\
\gamma&=&-2LT\left(G'(0)+\frac{1}{2}G'''(0)T+\frac{1}{12}G^{(5)}(0)T^2\right)
\end{eqnarray}

In the case of $SO(6)$ we again assume that the solution is reduceable
to $SO(4)\simeq SU(2)\times SU(2)$, with each $SU(2)$ having one cut of
roots.  Hence we have that $G(x)=G_+(x)+G_-(x)$, where $G_{\pm}(x)$ is
a single cut solution.  For such a rational solution the contribution
to the three-loop term is $\gamma^{(3)}=\gamma^{(3)}_++\gamma^{(3)}_-$ 
where using (\ref{3gamma}) and results in the appendix
\begin{equation}
\gamma^{(3)}_\pm=
\frac{\lambda^3s_\pm(n_\pm-s_\pm)(n^2_\pm-3s_\pm(n_\pm-s_\pm))(2s_\pm-n_\pm)^2}{16L^5}+{\rm O}(\lambda^4/L^7)
\end{equation}
If we now set $s_\pm=\pm\alpha n/2$ and $n_\pm=\pm n$, then 
\begin{equation}\label{gamma3}
\gamma^{(3)}=
\frac{\lambda^3n^6\alpha(2-\alpha)(4-3\alpha(2-\alpha))(1-\alpha)^2}{128 L^5}.
\end{equation}
The equation in (\ref{gamma3}) can be written more compactly in terms
of $\kappa=(1-\alpha)^2$, 
\begin{equation}\label{gamma3k}
\gamma=\frac{\lambda n^2(1-\kappa)}{4L}-\frac{\lambda^2n^4(1-\kappa)(1+3\kappa))}{64L^3}+\frac{\lambda^3n^6(1-\kappa)(1+3\kappa)\kappa}{128 L^5}
+{\rm O}(\lambda^4/L^7),
\end{equation}
where we have included the one and two-loop terms  for completeness.
In the appendix, the contribution to the energy at order $\lambda^3/L^5$
is computed for the semiclassical string by doing third order perturbation theory.
In terms of $\kappa$, this is 
\begin{equation}
E^{(3)}=\frac{\lambda^3n^6(1-\kappa)(1-3\kappa)(1-5\kappa)}{256 L^5},
\end{equation}
and so does not match the corresponding term in (\ref{gamma3k}).  
The difference between the terms is
\begin{equation}
E^{(3)}-\gamma^{(3)}=\frac{\lambda^3n^6(1-\kappa)(1-\kappa)(1-9\kappa)}{256L^5}
\end{equation}
The terms match at $\kappa=1$ as expected, since this is the BMN limit.
But curiously they also match at $\kappa=1/9$ which corresponds to $\alpha=2/3$
and so $J=L/3$.  

In the $(J',J',J)$ $SU(3)$ case the equation for the resolvent $W(x)$ in
(\ref{Weq}) is modified to
\begin{equation}\label{Weq2}
W(x+i0)+W(x-i0)+W(-x)=\frac{1}{x}+\frac{2T}{x^3}+{6T^2}{x^5}-2\pi n\qquad x\in\CC_+.
\end{equation}
and so $W_r(x)$ is
\begin{equation}
W_r(x)=-\frac{1}{3x}-\frac{2T}{3x^3}-\frac{2T^2}{x^5}-2\pi n.
\end{equation}
Equations (\ref{whom}), (\ref{req}) and (\ref{gxeq}) still apply,
although $r(x)$ and $s(x)$ need to be adjusted  since the asymptotic
behavior of $w(x)$ is modified to
\begin{equation}\label{wasymp2}
w(x)\approx-\frac{1}{3x}-\frac{2T}{x^3}-\frac{2T^2}{x^5}\qquad\quad x\to0.
\end{equation}
The functions $r(x)$, $s(x)$ and the corresponding solution
$g(x)$ up to order $T^2$ are given in the appendix.
The resulting anomalous dimension is
\begin{equation}\label{gamma3su3}
\gamma=-2TL\left(G'(0)+\frac{T}{2}G'''(0)+\frac{T^2}{12}G^{(5)}(0)\right)
=\frac{n^2\alpha\lambda }{2L}\left(1-
\frac{n^2\lambda}{4L^2}+\frac{n^4\lambda^2(1-\alpha^2)}{8L^4}\right)
\end{equation}
Notice that there is no three loop contribution if $\alpha=1$.  This
corresponds to $J=0$ where
the $SU(3)$ chain reduces to the $SU(2)$ 
chain, whose null three loop term was previously computed  
\cite{Serban:2004jf}.

Taking the expansion for the Frolov-Tsytlin semiclassical string
one loop higher, we find \cite{Frolov:2003qc}
\begin{equation}\label{Eeq}
E=L+\frac{n^2\alpha\lambda }{2L}\left(1-
\frac{n^2\lambda}{4L^2}+\frac{n^4\lambda^2(1-2\alpha+2\alpha^2)}{8L^4}\right)
+{\rm O}(\lambda/L^7)\end{equation}
As is now becoming quite familiar, the third order terms in (\ref{gamma3su3})
and (\ref{Eeq}) do not match.  But once again, they do match for the special
case of $\alpha=2/3$ ($J=L/3$), which in the circular string corresponds to
all three $R$-charges being equal.  In terms of the $SU(3)$ subgroup of
the $SU(4)$ $R$-symmetry group, this state is an $SU(3)$ singlet.
As in the pulsating string case, it is not clear if this a happy coincidence
or a clue toward resolving the divergence of the gauge and string expansions.

\sectiono{$SO(6)$ Fluctuations}

This section is somewhat outside the main development of this paper.
For rational $SU(2)$ 
solutions it is quite easy to find the spectrum of fluctuations
about the one-loop 
semiclassical solution.  We first present a method applicable to
$SU(2)$ solutions and then apply it to the $SO(6)$ case.

The idea is to remove one Bethe root from the cut, essentially placing
it on a different branch.  Assuming a one cut Bethe solution with
momentum $2\pi s$ and branch number $n$, then if one root is moved onto
a branch with branch number $n_1=n+\Delta n$, then the position of the root $x_1$ in the rescaled
complex plane satisfies the equation
\begin{equation}
\frac{1}{x_1}-2\pi n_1=2\int_\CC \frac{dx'\sigma(x')}{x_1-x'}=2G(x_1),
\end{equation}
where $G(x)$ has the form in (\ref{Grat}).
Hence $x_1$ is given by
\begin{eqnarray}\label{flucpos}
x_1&=&\frac{1}{2\pi}\frac{(2s-n)\pm\sqrt{(\Delta n)^2-4s(n-s)}}{(\Delta n)^2-n^2}
\qquad \Delta n\ne n
\nonumber\\
x_1&=&\frac{1}{4\pi(n-2s)}\qquad\qquad\qquad\qquad\qquad \Delta n=n
.
\end{eqnarray}

To compute the change in the anomalous dimension due to this movement of
the root, we need to compute the back reaction on the cut 
\cite{Beisert:2003xu,Freyhult:2004iq}.
The effect of this root is modify the integral equation for the cut to
\begin{equation}
\frac{1}{x}-2\pi n=2\pint_\CC\frac{dx'\sigma(x')}{x-x'}+\frac{2/L}{x-x_1}
\qquad x\in\CC,
\end{equation}
where $x_1$ is given in (\ref{flucpos}).
Hence, we need to solve the Riemann-Hilbert problem
\begin{equation}
G(x+i0)+G(x-i0)=\frac{1}{x}-2\pi n-\frac{2/L}{x-x_1}.
\end{equation}
The general form for $G(x)$ is
\begin{equation}
G(x)=\frac{1}{2x}-\frac{1/L}{x-x_1}+\frac{1}{2}\left(\frac{1}{x}+
\frac{b}{x-x_1}\right)\sqrt{A^2x^2+Bx+1}-\pi n.
\end{equation}
In order to cancel the pole at $x=x_1$, we have that 
\begin{equation}
\frac{b}{2}\sqrt{A^2x^2+Bx+1}=\frac{1}{L}.
\end{equation}
Approximating $A$ and $B$ by their values in (\ref{Grat}), we obtain
\begin{equation}
b\approx-\frac{1}{\pi L \Delta n x_1}.
\end{equation}
In order to have the correct asymptotic behavior for $G(x)$, we
also have
\begin{eqnarray}
A&\approx&2\pi n(1-b)\nonumber\\
B&\approx&4\pi(2s-n)(1-2b)+\frac{8\pi n^2}{L\Delta n}.
\end{eqnarray}

The momentum is given by
\begin{equation}
\frac{1}{x_1}-G(0)=2\pi s +\frac{2\pi\Delta n}{L}
\end{equation}
and the one-loop anomalous dimension is
\begin{equation}
\gamma=\frac{\lambda}{8\pi^2}\left(\frac{1}{L^2x_1^2}-\frac{G'(0)}{L}\right).
\end{equation}
Hence the change in the anomalous dimension due to the fluctuation is
\begin{equation}\label{gammafluc}
\Delta\gamma=\frac{\lambda\Delta n}{2L^2}\left(\sqrt{(\Delta n)^2-4s(n-s)}-2(2s-n)\right).
\end{equation}
This matches the form in \cite{Frolov:2003qc,Arutyunov:2003za}.

Examining (\ref{gammafluc}) we see that if $(\Delta n)^2<4s(n-s)$ then
$\Delta\gamma$ is complex, signaling an instability.  If we look
at (\ref{flucpos}), we see that this corresponds to $x_1$ moving
off the real axis, an effect also seen in \cite{Freyhult:2004iq}.  
A physical Bethe state must have Bethe roots that are 
real, or are in conjugate pairs.  Thus the instability for the
quantum state can be thought
of as a Bethe root pushed off into the forbidden region.  However,
an allowed state can still have such a root if it also has the conjugate.

For circular strings,
one has that $s=m>0$ and that $n\ge 2m$.  Hence circular strings always
have an instability for the $\Delta n=\pm 1$ mode since $s(n-s)\ge1$.

For pulsating strings, if we have $s_+=-s_-$ and $n_+=-n_-=1$, then there
is no instability.  The string with $\alpha=1$ is just barely stable,
since here one finds that the $\Delta n=1$ mode is massless.  
Note that the physical states still have to satisfy the momentum condition.
Hence a state requires at least two fluctuations, with the sum over all
$\Delta n$ being zero.

\sectiono{Discussion}

In this paper we have exploited the ``closed sector reduction'' for certain
types of long
coherent operators in order
to compute  higher loop contributions to their anomalous
dimensions.  At the two-loop level we find agreement with string predictions
for the anomalous dimensions.  It should be straightforward to
verify that all higher charges match as well, using arguments and
procedures given in 
\cite{Arutyunov:2003rg,Engquist:2004bx,Kazakov:2004qf,Arutyunov:2004xy}.
It would also be interesting to consider the higher loop contributions
for more general operators in the $SU(3)$ sector, for example, those
considered in \cite{Kristjansen:2004ei}. 

At three-loops there is disagreement except for the special cases.  
The Inozemtsev model
 does not have perturbative BMN scaling at four loops \cite{Serban:2004jf} 
and so
there can be no way 
to match the gauge and string theory predictions at
the four-loop  level with this model. 
However, very recently a new model was proposed for the
$SU(2)$ chain which
assumes that BMN scaling holds to all orders in perturbation theory
\cite{Beisert:2004hm}.  One could modify the Bethe equations for the $SU(3)$ 
and $SO(6)$ chains as outlined here.  We
would not expect agreement for general operators at four loops or higher,
but it would be interesting to see if the $(J,J,J)$ state agrees at this
level.

Resolving the mystery of three loops is a crucial problem in our
understanding of the AdS/CFT correspondence 
\cite{Maldacena:1998re,Gubser:1998bc,Witten:1998qj}.  If the correspondence
is correct, then presumably this is an issue of strong
versus weak coupling and there are contributions that do not appear in
perturbation theory but do contribute to the classical string, or
vice versa.  In the examples provided
here, while there is disagreement at three loops, the difference between
the predictions
are simple rational expressions.  Quantities that are rational are often
computable, giving one hope that ultimately a resolution can be found.

\bigskip

\noindent {\bf Acknowledgments}:
I thank K. Zarembo for many helpful discussions.  This research was
supported in part by Vetenskapsr\aa det
and by DOE contract \#DE-FC02-94ER40818.

\appendix

\sectiono{Some results for three loops}

\subsection{Single cut $SU(2)$}

At three loops, the Riemann Hilbert problem becomes
\begin{equation}
G(x+i0)+G(x-i0)=\frac{1}{x}+\frac{2T}{x^3}+\frac{6T^2}{x^5}-2\pi n\qquad x\in\CC.
\end{equation}
With only one cut on branch $n$, the general form of $G(x)$
is
\begin{equation}
G(x)=\frac{1}{2x}+\frac{T}{x^3}+\frac{3T^2}{x^5}
+\frac{1}{2}\left(\frac{a_1}{x}+\frac{a_2}{x^2}+\frac{a_3}{x^3}+
\frac{a_4}{x^4}+\frac{3T^2}{x^5}\right)\sqrt{A^2x^2+Bx+1}-\pi n.
\end{equation}
We also set the total momentum to $2\pi s$ and so $G(x)$ is asymptotically
\begin{equation}
G(x)\approx \frac{s/n}{x}\qquad x\to 0,
\end{equation}
since the number of roots is $s/n$.  The $a_i$ can be determined in terms
of $A$ and $B$ by canceling the singularities at $x=0$.  The asymptotic
conditions then give us the two further equations
\begin{eqnarray}\label{ABeqs}
a_1{A}&=&2\pi n\nonumber\\
1+\frac{a_1B}{2A}+a_2A&=&\frac{s}{n}\,.
\end{eqnarray}
The results are
\begin{eqnarray}
a_1&=&1-\left(A^2-\frac{3}{4}B^2\right)T+\left(\frac{9}{4}A^4-\frac{45}{8}B^2A^2+\frac{105}{64}B^4\right)T^2\nonumber\\
a_2&=&-BT+\left(\frac{9}{2}BA^2-\frac{15}{8}B^3\right)T^2\nonumber\\
a_3&=&2T-\left(A^2-\frac{9}{4}B^2\right)T^2\nonumber\\
a_4&=&-3BT^2\,.
\end{eqnarray}
These equations and the conditions in (\ref{ABeqs}) lead to the following
approximations for the variables:
\begin{eqnarray}
A&=&2\pi n\Big(1-2T(2\pi )^2(n^2-6s(n-s))\\
\nonumber
&&\qquad\qquad\qquad\qquad+6T^2(2\pi)^4(n^4-16n^3s+66n^2s^2-100n^s+50s^4)\Big)+{\rm O}(T^3)\nonumber\\
B&=&4\pi(2s-n)\Big(1-2T(2\pi)^2(n^2-12s(n-s))\\
\nonumber
&&\qquad\qquad\qquad\qquad+6T^2(2\pi)^4(n^4-30n^3s+154n^2s^2-248ns^3+124s^4)\Big)
+{\rm O}(T^3)
\end{eqnarray}
\begin{eqnarray}
a_1&=&1+2T(2\pi)^2(n^2-6s(n-s))-2T^2(2\pi)^4(n^4-24n^3s+102n^2s^2-156ns^3+78s^4)
+{\rm O}(T^3)\nonumber\\
a_2&=&-2T(2\pi)(2s-n)-2T^2(2\pi)^3(2s-n)(n^2-6s(n-s))
+{\rm O}(T^3)\nonumber\\
a_3&=&2T+6T^2(2\pi)^2(n^2-6s(n-s))
+{\rm O}(T^3)\nonumber\\
a_4&=&-6T^2(2\pi)(2s-n)
+{\rm O}(T^3)
\end{eqnarray}
Expanding $G(x)$ about $x=0$ we find
\begin{eqnarray}
G'(0)&=&-(2\pi)^2s(n-s)
+ 
2T(2\pi)^4s(n-s)(n-3s)(2n-3s)
\nonumber\\
&&\qquad\qquad-2T^2(2\pi)^6s(n-s)(8n^4  
-3s(n-s)(29n^2-74s(n-s)))+{\rm O}(T^3)\nonumber\\
\frac{1}{3!}G'''(0)&=&-(2\pi)^4s(n-s)(n^2-5s(n-s))\nonumber\\
&&\qquad\qquad+4T(2\pi)^6s(n-s)(2n^4 
-5s(n-s)(5n^2-14s(n-s)))+{\rm O}(T^2)\nonumber\\
\frac{1}{5!}G^{(5)}(0)&=&-(2\pi)^6s(n-s)(n^4-14s(n-s)(n^2-3s(n-s)))+{\rm O(T)}
\end{eqnarray}

\subsection{Terms for $SU(3)$}

The functions $r(x)$ and $s(x)$ are determined by matching the singularities
in equations (\ref{req}), (\ref{wasymp}), (\ref{wasymp2}) and (\ref{gxeq}).
The results are
\begin{eqnarray}
r(x)&=&\frac{1}{3x^2}+(2\pi n)^2+\left(\frac{4}{3x^4}+\frac{2(2\pi n)^2\alpha}{x^2}\right)T+\left(\frac{16}{3x^6}+\frac{6(2\pi n)^2\alpha}{x^4}-\frac{2(2\pi n)^4\alpha}{x^2}\right)T^2+{\rm O}(T^3)\,,\nonumber\\
s(x)&=&\frac{2}{27x^3}-\frac{(2\pi n)^2(2-3\alpha)}{3x}
+\left(\frac{4}{9x^5}-\frac{4(2\pi n)^2(1-2\alpha)}{3x^3}\right)T
\nonumber\\
&&\qquad\qquad+\left(\frac{20}{9x^7}-\frac{4(2\pi n)^2(3-7\alpha)}{3x^5}+\frac{2(2\pi n)^4\alpha(2-3\alpha)}{3x^3}\right)T^2+{\rm O}(T^3)\,.
\end{eqnarray}
We then find for $g(x)$ 
\begin{eqnarray}
g(x)&=&\left(-\frac{2}{3x}-\alpha(2\pi n)^2 x-\alpha(1-2\alpha)(2\pi n)^4x^3-
\alpha(1-6\alpha+7\alpha^2)(2\pi n)^6x^5 + {\rm O(x^7)}..\right)\nonumber\\
&&\qquad\qquad+\left(-\frac{4}{3x^3}+2\alpha(2-3\alpha)(2\pi n)^4x+8\alpha(1-5\alpha(1-\alpha))(2\pi n)^6x^3+{\rm O}(x^5)\right)T\nonumber\\
&&\qquad\qquad+\left(-\frac{4}{x^5}-4\alpha(4-15\alpha+12\alpha^2)(2\pi n)^6x+{\rm O}(x^3)\right)T^2 +{\rm O}(T^3)
\end{eqnarray}

\subsection{Third order perturbation theory}

The general expression for the third order correction to the energy from
a perturbation $\HH'$ to a Hamiltonian is
\begin{equation}\label{3rdorder}
\varepsilon^{(3)}=\sum_{\ell'\ne\ell,\ell''\ne\ell}
\frac{\langle\ell|\HH'|\ell'\rangle\langle\ell'|\HH'|\ell''\rangle
\langle\ell''|\HH'|\ell\rangle}{(\ve_\ell-\ve_{\ell'})(\ve_\ell-\ve_{\ell''})}
-\sum_{\ell'\ne\ell}\frac{\langle\ell|\HH'|\ell'\rangle\langle\ell'|\HH'|\ell\rangle\langle|\ell|\HH'|\ell\rangle}{(\ve_\ell-\ve_{\ell'})^2}\,.
\end{equation}
In our case $\HH'=\lambda n^2(1-w)$ and so
 $\langle\ell'|\HH'|\ell\rangle\ne0$ only if $\ell'=\ell\pm1$, $\ell'=\ell$. Hence, the expression in (\ref{3rdorder}) reduces to
\begin{equation}
\varepsilon^{(3)}_\ell=\sum_{\ell'=\ell\pm1}
|\frac{\langle\ell|\HH'|\ell'\rangle|^2}{(\ve_\ell-\ve_{\ell'})^2}(\langle\ell'|\HH'|\ell'\rangle-\langle\ell|\HH'|\ell\rangle)
\end{equation}
If we define
$\phi(\ell+1/2)=\langle\ell|\HH'|\ell+1\rangle$ and
$\psi(\ell)=\langle\ell|\HH'|\ell\rangle$, then for large $\ell$, we
can approximate $\ve^{(3)}_\ell$ by
\begin{equation}
\ve^{(3)}_\ell=\frac{\phi^2(\ell)\psi''(\ell)}{(\ve'(\ell))^2}+2\frac{\phi(\ell)\phi'(\ell)\psi'(\ell)}{(\ve'(\ell))^2}-2\frac{\ve''(\ell)\phi^2(\ell)\psi'(\ell)}{(\ve'(\ell))^3},
\end{equation}
where from (\ref{1storder}) and (\ref{2ndorder}), we can approximate
\begin{equation}
\psi(\ell)=2\phi(\ell)=\lambda n^2\frac{\ell^2-j^2}{2\ell^2}\qquad\ve(\ell)=\ell^2,
\end{equation}
Hence we find that 
\begin{equation}
\ve^{(3)}_\ell=(\lambda n^2)^3\frac{\kappa(1-\kappa)(9\alpha-5)}{64L^4}
\end{equation}
where $\kappa=j^2/\ell^2$.  Hence, up to three loops, the energy squared
of the string state is
\begin{equation}
E^2=L^2+\frac{\lambda n^2(1-\kappa)}{2}+\frac{(\lambda n^2)^2(1-\kappa)(1-5\kappa)}{32L^2}+\frac{(\lambda n^2)^3\kappa(1-\kappa)(9\kappa-5)}{64L^4} 
+{\rm O}(\lambda^4/L^6)
\end{equation}
and so $E$ is
\begin{equation}
E=L+\frac{\lambda n^2(1-\kappa)}{4L}-\frac{(\lambda n^2)^2(1-\kappa)(1+3\kappa)}{64L^3}+\frac{(\lambda n^2)^3(1-\kappa)(1-3\kappa)(1-5\kappa)}{256L^5}\, .
\end{equation}

\end{document}